\documentclass[prb,aps,twocolumn,superscriptaddress]{revtex4}

\def\bfl{\begin{flushleft}}
\def\efl{\end{flushleft}}
\def\bfr{\begin{flushright}}
\def\efr{\end{flushright}}
\def\bc{\begin{center}}
\def\ec{\end{center}}
\def\be{\begin{equation}}
\def\ee{\end{equation}}
\def\ba{\begin{eqnarray}}
\def\ea{\end{eqnarray}}
\def\baa#1{\begin{array}{#1}}
\def\eaa{\end{array}}
\def\bw{\begin{widetext}}
\def\ew{\end{widetext}}

\def\text#1{\mbox{#1}}

\begin{document}

\title{Comment on "Fe valency-induced effect on the magnetic and electrotransport properties in Nd$_{0.67}$Sr$_{0.33}$Mn$_{1-x}$Fe$_x$O$_3$ polycrystalline system" [Appl. Phys. Lett. 82, 1721 (2003)]}

\author{
Andrew Das Arulsamy \thanks{$E-mail: sadwerdna@hotmail.com}}

\affiliation{
Department of Physics, National University of Singapore, 2 Science Drive 3,
Singapore 117542, Singapore}

\affiliation{
Condensed Matter Group, No. 22, Jln Melur 14, Tmn Melur, Ampang,
Selangor DE, Malaysia}

\maketitle

\narrowtext

The authors, YL Chang and CK Ong (CO)~\cite{CO1} recently have reported several interesting observations on magnetic and electrotransport properties of Nd$_{0.67}$Sr$_{0.33}$Mn$_{1-x}$Fe$_x$O$_3$. CO have carried out XPS, XRD, magnetic and electrotransport measurements on the said compound to illustrate its physical properties and its applications. Firstly, CO claimed that their experimental data on XPS indicate that Mn's average valence state decreases with Fe content. This implies that Fe doping produces more Mn$^{3+}$ than Mn$^{4+}$ relatively. However, one can easily notice that the intensity peaks of Mn 2$p$ for both Mn$^{3+}$ and Mn$^{4+}$ were decreased identically with Fe content as shown in Fig. 3a of Ref.\cite{CO1}. These peaks do not show Mn$^{4+}$ (MnO$_2$) intensity peak reduces more rapidly than Mn$^{3+}$ (Mn$_2$O$_3$) peak. As such, the argument of Mn$^{4+}$ $\to$ Mn$^{3+}$ with Fe content and subsequently the conclusion of reduction in Mn's average valence state is incorrect. It is important to realize that Gutierrez {\it et al}.~\cite{guti2} have concluded with titration measurements that the ratio of Mn$^{3+}$/Mn$^{4+}$ increases (Mn's average valence state decreases) with Fe$^{3+}$ doping in La$_{0.7}$Pb$_{0.3}$Mn$_{1-y}$Fe$_y$O$_3$ for $y$ = 0 $\to$ 0.3. Note here that Fe$^{3+}$ is maintained and the concentration of Fe$^{4+}$ does not increase with Fe content in this observation. In addition, increasing Fe$^{3+}$ content is found to reduce hopping electrons that suppresses double exchange mechanism and finally weakening of ferromagnetism and metallic properties~\cite{guti2}. In contrast, the XPS data from CO neither prove nor justify their own claim on the reduction of Mn's valence state. It is worth noting that CO not only claim to observe the decrement of Mn's valence state with Fe content in accordance with Gutierrez {\it et al}., but CO also found that Fe doping increases the Fe's valence state for $x$ $>$ 0.1 violating Gutierrez {\it et al}.'s observation. Simply put, Gutierrez {\it et al}. were logically correct to claim that average Mn's valence state reduces with Fe$^{3+}$ doping so as to compensate the larger size of Fe$^{3+}$. This is an important scenario that maintains the purity of the crystal structure with doping. According to CO however, ionic sizes of Mn and Fe increases and reduces respectively for $x$ $>$ 0.1, which is in direct opposition to Ref.~\cite{guti2} in order to maintain crystal stability. It is absurd to surmise that smaller ions substitute larger ones. On the contrary, instead of substituting larger Mn ions, Fe ions will likely to form interstitial defects in crystals. I.e., Fe$^{3+}$'s (0.7825 {\AA}) subsititution into Mn$^{3+}$(0.7825 {\AA})/Mn$^{4+}$(0.67 {\AA}) system for $x$ $>$ 0.1 enhances Mn$^{4+}$ $\to$ Mn$^{3+}$ to compensate the larger size of Fe$^{3+}$. Unlike Fe$^{3+}$, substitution of Fe$^{3+}$/Fe$^{4+}$(0.725 {\AA}) into Mn$^{3+}$/Mn$^{4+}$ system for $x$ $>$ 0.1  will enhance both Mn$^{4+}$ $\to$ Mn$^{3+}$ and Mn$^{3+}$ $\to$ Mn$^{4+}$. The former {\it decrement} of Mn's average valence state is due to compensation of larger size, Fe$^{3+}$(0.7825 {\AA}) and Fe$^{4+}$(0.725 {\AA}) compared to Mn$^{4+}$(0.67 {\AA}). While the latter {\it increment} of Mn's average valence state is due to the compensation of smaller size, Fe$^{4+}$(0.725 {\AA}) compared to Mn$^{3+}$(0.7825 {\AA}). Therefore, apart from the possibility of forming interstitial defects, the latter increment also competes with the former decrement to suppress the overall decrement of Mn's average valence state for $x$ $>$ 0.1 instead of decreasing it systematically. As a consequence, CO's XPS and XRD data on Fe are somewhat doubtful. 

Secondly, CO have also misinterpreted that resistivity increases due to AFM interaction that reduces the magnetization. Actually, the magnitude of resistivity increases in a {\it continuous way with temperature} ($T$) due to reduced magnetization complying with reduction in Curie temperature~\cite{guti2} ($T_C$). Reduction in magnetization does not shift the whole curve upward with Fe content as incorrectly concluded by CO. This reduction in magnetization does not imply increasing $\rho(T)$ with Fe doping in any way. 

Subsequently, CO argued that "some of the iron ions begin to transit to tetravalence state with only singly unoccupied $e_g$ orbital in the high spin state" hence it gives rise to electrical conduction with Fe$^{4+}$ as $x$ increases above 0.1. Parallel to this, CO also conclude "the electrical conductivity in the system is observed to increase with more Fe$^{4+}$, which enables greater transfer of electrons". Note also that CO already claimed from XPS data on Fe stating that Fe$^{4+}$ increases with Fe doping for $x$ $>$ 0.1 where Fe 2$p$ spectra resemble 2$p_{3/2}$ peaks of Nd$_{0.67}$Sr$_{0.33}$FeO$_3$ (NSFO). I.e., for $x$ $>$ 0.1 both Nd$_{0.67}$Sr$_{0.33}$MnO$_3$ (NSMO) and NSFO determine the ground state~\cite{CO1}. In contrast, from $\rho(T)$ data (Fig. 4 of Ref.\cite{CO1}), $\rho(T)$ keeps on increasing systematically with Fe content even for $x$ = 0.3 $\to$ 0.4, which is in contradiction with their own statement mentioned above (conductivity is enhanced with Fe$^{4+}$ for $x$ $>$ 0.1 due to superexchange mechanism). Now assuming CO were right, if electrical conduction is really enhanced by Fe$^{4+}$ for $x$ = 1 compared to $x$ = 0.3 and 0.4 as claimed by CO, then the rate of $\rho$ increasing with $T$ for $x$ = 1 should not be more rapid than samples with $x$ = 0.3 and 0.4 since the so-called greater transfer of hopping $e_g$ electrons, which is proportional to Fe$^{4+}$ should not give rapid increase of $\rho$ with $T$. On the contrary, inconsistency can be seen in $\rho(T)$ curve (Fig. 4 of Ref.\cite{CO1}) for $x$ = 1 that shoots up more rapidly than samples with $x$ = 0.3 and 0.4. Consequently, the whole curve of $\rho(T)$ at $x$ = 1 shifted downward below $x$ = 0.4 is obviously not due to easy hopping of $e_g$ electrons. As a matter of fact, the transfer of $e_g$ electrons determines the rate of $\rho$ increases or decreases with $T$ and also the suppression or enhancement of ferromagnetism respectively with doping. The latter incompatibility makes CO's XPS and XRD data on Fe even more doubtful. Briefly, increment of Fe$^{4+}$ with doping for $x$ $>$ 0.1 as claimed to be observed by CO is not compatible with systematic decrement of Mn's average valence state and upward or downward shifting of $\rho(T)$ curves. However, it is likely that Mn's average valence state decreases with Fe$^{3+}$ doping for $x$ = 0 $\to$ 0.4 that suppressed double exchange mechanism due to lack of Mn$^{4+}$ in accordance with suppressed magnetization and $T_C$ (from electrotransport measurements). The correct mechanism that could predict and explain the increasing $\rho(T)$ for $x$ = 0 $\to$ 0.4 and an unexpected downward shift of $\rho(T)$ for $x$ = 1.0 are incomplete presently. CO should realize that their assumption of increasing Fe$^{4+}$ for $x$ $>$ 0.1 that gives rise to superexchange mechanism does not explain the above $\rho(T)$ observations at all. All the values of ionic sizes reported here were taken from Ref.~\cite{web3}.         

In conclusion, the abstract and conclusions as appear in Ref.~\cite{CO1} contain several scientifically flawed claims with possible error(s) in CO's XPS and XRD data that will eventually lead to severe confusion to the readers in this field and also will lead to further misunderstandings and misinterpretations in future.
   
ADA would like to thank the National University of Singapore for the financial aid. The author is grateful to A. Innasimuthu, I. Sebastiammal, A. Das Anthony and Cecily Arokiam for the partial financial assistances.

\end{document}